# Isolated Flat Bands and Physics of Mixed Dimensions in a 2D Covalent Organic Framework

*Juefan Wang[1,2], Su Ying Quek[1,2, *]*

[1]Department of Physics, National University of Singapore, 2 Science Drive 3, 117551, Singapore

[2]Centre for Advanced 2D Materials, National University of Singapore, Block S14, Level 6, 6 Science Drive 2, 117546, Singapore

**Abstract**

We demonstrate that it is possible to rationally incorporate both an isolated flat band, and the physics of zero dimensions (0D), one dimension (1D), and two dimensions (2D) in a single 2D material. Such unique electronic properties are present in a recently synthesized 2D covalent organic framework (COF), where "I"-shaped building blocks and "T"-shaped connectors result in quasi-1D chains that are linked by quasi-0D bridge units arranged in a stable 2D lattice. The lowest unoccupied conduction band is an isolated flat band, and electron-doping gives rise to novel quantum phenomena, such as magnetism and Mott insulating phases. The highest occupied valence band arises from wave functions in the quasi-1D chains. Examples of mixed dimensional physics are illustrated in this system. The strong electron-hole asymmetry in this material results in a large Seebeck coefficient, while the quasi-1D nature of the chains leads to linear dichroism, in conjunction with strongly bound 2D excitons. We elucidate strategies to design and optimize 2D COFs to host both isolated flat bands and quantum-confined 1D subsystems. The properties of the 2D COF discussed here provide a taste of the intriguing possibilities in this open research field.

**Introduction**

The dimensionality of a material has profound implications on its properties. The reduced dielectric screening in one dimension (1D) results in large exciton binding energies in semiconducting nanotubes and nanoribbons,[1] and even leads to bound excitons in metallic carbon nanotubes.[2] Depolarization effects[3] in these 1D systems implies that a strong optical response is observed only for light polarized along the tube or ribbon axis. Strong electron correlation in 1D systems can also lead to exotic phenomena such as spin-charge separation in Luttinger liquids, and the emergence of correlated-electron insulators.[4]

Physics in two dimensions (2D) is also fundamentally interesting. The properties of 2D materials can be readily tuned by gating and the choice of substrates. 2D materials are also flexible and can be stacked together to form heterostructures with new properties. Arguably, one of the most significant developments in 2D materials to date was the recent discovery that twisted bilayer graphene can become superconducting when the twist angle is about 1.1°.[5] Central to this novel behavior was the emergence of a flat band close to the Fermi level. This seminal discovery has revived interest in the physics of flat bands, and precipitated a materials search for flat bands in other twisted bilayers.[6] The flat bands in twisted bilayers have been attributed to the weak coupling between states localized at regions with specific stacking registries[7,8] when the twist angle is sufficiently small. These flat bands can potentially host a wide array of exotic quantum states, such as superconductivity,[5] ferromagnetism,[9] spin liquid states[10] and Mott-Hubbard insulators.[11]

The excitement of probing flat band physics has resulted in a significant amount of attention on the properties of twisted bilayers. However, experimentally, obtaining the small twist angles can be challenging as the bilayers prefer the zero twist angle stacking configuration. Instead of a top-down approach, it is also possible to obtain frontier electronic states that are flat bands, through bottom-up synthesis of covalent organic frameworks (COFs)[12-15] or metal organic frameworks (MOFs)[16-19] with specific lattices. However, in most cases, these flat bands are degenerate or nearly degenerate with a dispersive band,[12-19] which can result in mixing between bands in the presence of additional Coulomb interactions.[20] A large number of COFs and MOFs with flat bands can be described by Kagome lattices.[12-14, 18, 19] In a Kagome lattice, a flat band arises from destructive interference in the wavefunction.[20-23] Mathematically, it can be proven that such a flat band must be degenerate with a dispersive band,[20] and any effects that lift this degeneracy will result in increased dispersion in the flat band.[20]

COFs are organic materials in which molecular building blocks are linked by strong covalent bonds, forming an extended periodic network with ordered pores. COFs are an active topic of research in the chemistry community, with much research primarily focused on their applications to important problems in energy storage, healthcare and environmental protection.[24, 25] At the same time, one beauty of COFs is that scientists can design from bottom-up its structural topology by choosing the molecular building blocks and linker units.[26] In this work, we show that this bottom-up structural design also enables one to engineer the electronic structure of COFs to host an interesting interplay of mixed-dimensional physics. We focus on a class of experimentally synthesized 2D-COFs with "I"-shaped building blocks that are linked together with "T"-shaped motifs,[27, 28] resulting in quasi-1D chains connected by quasi-zero dimensional (0D) bridge units in a periodic 2D lattice. We demonstrate that one of these COFs, 2D-CAP, naturally hosts the long sought-after isolated flat band, which constitutes the lowest unoccupied conduction band. Unlike the twisted bilayers and other COFs discussed above, 2D-CAP not only has flat bands, but also presents unique opportunities to study quasi-1D physics in conjunction with 2D physics and flat bands, within a single 2D sheet. For example, the quasi-1D nature of the chains leads to linear dichroism in conjunction with 2D excitons that are distributed among the 0D bridge sites. The highest occupied valence band wave function is localized primarily on the quasi-1D chains, and band dispersion along the direction of these chains, together with the flat conduction band, leads to strong electron-hole asymmetry and a large Seebeck coefficient. The isolated flat band can also result in novel quantum phases. Electron doping of the 2D-CAP leads to magnetism, and the on-site interaction at the bridge units is large, suggesting the possibility of a Mott-Hubbard insulating phase. We discuss strategies to design and optimize the properties of other 2D COFs that can also host isolated flat bands and physics of mixed dimensions.

**Methods**

Density functional theory (DFT) calculations were performed, with the exchange-correlation functional treated within the generalized gradient approximation,[29] as implemented in the plane-wave pseudopotential code, Quantum Espresso.[30] Spin polarization was included for calculations on doped materials, where a jellium neutralizing background charge is added. Grimme's D2 dispersion corrections were added to compute the exfoliation energy, using the following equation:[31]

$$E_{exf} = \frac{E_{iso} - E_{bulk}/2}{A} \qquad (1)$$

where $E_{exf}$ is the exfoliation energy per unit area, $E_{iso}$ is the energy per unit cell of the isolated monolayer, $E_{bulk}$ is the energy per unit cell of bulk 2D-CAP (eclipsed stacking configuration with two layers per unit cell). The SG15 Optimized Norm-Conserving Vanderbilt pseudopotentials (version 2.0.1) were used.[32] Plane-wave cutoffs of 80 Ry for the wave functions and 320 Ry for the charge density were sufficient to obtain converged results. A vacuum length of 12 Å was used to prevent interactions between the monolayers. Monkhost-Pack k-point meshes of 4 × 4 × 1 and 5 × 5 × 1 were used for the geometry optimization of monolayer 2D-CAP and CAP-1 unit cells, respectively. The convergence thresholds for the total energy and force were $10^{-7}$ Ry and $10^{-5}$ Ry/Bohr, respectively. For phonon calculations, the convergence threshold for self-consistency was $10^{-18}$ Ry. A 4 × 4 × 1 q-point mesh was used for the phonon dispersion of monolayer 2D-CAP unit cell.

To study the optical properties of monolayer 2D-CAP, GW-Bethe-Salpeter equation (BSE) calculations were performed using the BerkeleyGW code,[33] including self-energy effects and electron-hole interactions in the exciton energies. One-shot $G_0W_0$ calculations were performed with a dielectric matrix cutoff of 10

Ry, and a slab coulomb truncation scheme[34] to prevent interaction between periodic slabs. The Hybertsen-Louie generalized plasmon pole model[35] and static remainder approximation[36] were used in the GW calculation, and the Tamm-Dancoff approximation was used in the BSE calculation. The GW quasiparticle energies were converged to within 0.02 eV using a 4 × 4 × 1 k-point mesh and 1200 bands. The optical absorption spectra were converged with a 16 × 16 × 1 fine mesh.

The Seebeck coefficients were computed using the BoltTraP2 code[37] in the framework of semiclassical Boltzmann theory in the constant relaxation time approximation. The code used, as input, DFT eigenvalues computed using the plane wave code, VASP.[38] The DFT band structures from VASP and Quantum Espresso match closely. Further details are provided in the supporting information.

Tight binding calculations were also performed in this work, with details given in the section below.

**Results and Discussion**

Fig. 1(a-b) show the atomic structure of two recently synthesized 2D-COFs, CAP-1 (Fig. 1a) and 2D-CAP (Fig. 1b). The names, CAP-1 and 2D-CAP, are adopted from the experimental papers,[27, 28] and CAP stands for 'conjugated aromatic polymer'. The building blocks of these COFs are shaped like the upper-case letter "I", as shown in the unit cells depicted by parallelograms in the Figure. These building blocks are connected with "T"-shaped motifs. The design of these building blocks and linking motifs results in an array of quasi-0D bridge units arranged periodically in a 2D lattice, and linked by quasi-1D chains that extend infinitely in the y-direction (Fig. 1(a-b)). The on-site energy of the bridge units can be modified through functional groups, and in this case, nitrogen atoms replace carbon atoms at specific locations in the fused benzene rings. The bridge unit is longer for 2D-CAP than CAP-1, which controls the extent of coupling between the bridge units and the chains.

These 2D COFs were synthesized in bulk form from three-dimensional crystals in which the molecular building blocks are arranged in such a way that steric hindrance favours the formation of the layered 2D

COF structure when the molecular crystal is heated.[27, 28] The computed exfoliation energies are 12.4 meV/Å$^2$ and 16.6 meV/Å$^2$ for CAP-1 and 2D-CAP, respectively. The small values of these exfoliation energies indicate that monolayers of CAP-1 and 2D-CAP can be readily exfoliated.[39] Geometry optimization shows that these monolayer COFs exhibit a small amount of corrugation in the out-of-plane direction (Fig. 1(a-b)).

In previous work, researchers had focused on the applications of 2D-CAP and CAP-1 as organic electrodes,[27, 28] while the electronic structure properties were largely unreported. Fig. 1(c-d) show the band structures for these materials, computed using DFT calculations. Both systems exhibit strong electron-hole asymmetry. The highest occupied valence band has significant dispersion, except in the direction corresponding to $q_{bridge}$, while the lowest unoccupied conduction band is nearly flat in CAP-1, and very flat in 2D-CAP. The flat bands are also isolated in energy from other bands. This is in contrast to the nearly flat bands reported in the majority of other 2D-COFs, where the flat bands are degenerate or nearly degenerate with dispersive bands.[12-15] The isolated flat band for 2D-CAP is present even if the 2D-CAP monolayer is restricted to be completely planar, instead of taking its more stable, slightly corrugated, structure.

The strong electron-hole asymmetry can be understood from the wave function character of the valence band maximum (VBM) and conduction band minimum (CBM), as shown in Fig. 2. Both VBM and CBM wave functions have π–character. The 2D-CAP CBM is predominantly localized on the bridge sites, while the 2D-CAP VBM is predominantly localized on the quasi-1D chains. Similar observations are made for CAP-1. However, for CAP-1, the CBM wave function has more weight on the chains, explaining the slight dispersion of the conduction band in Fig. 1(c). Negligible interaction among the bridge sites results in the lowest unoccupied conduction band being very flat, especially for the 2D-CAP. On the other hand, electronic coupling along the quasi-1D chains leads to significant dispersion in the highest occupied valence bands, especially in the direction corresponding to $q_{chain}$.

Comparing CAP-1 and 2D-CAP, one can understand a few design principles in achieving the highly sought-after isolated flat band in 2D-COFs. It is instructive to adopt a tight-binding model to describe the key bands in question, and we have used an effective Kagome-like lattice as shown schematically in Fig. 1(a-b). Kagome-like lattices have been used in the literature for related COF materials.[12-14] The choice of lattice to model the 2D-COFs here is not unique (see Fig. S1 for results using a Kagome-honeycomb lattice model). However, we have chosen a Kagome-like lattice because it is sufficient to describe the bands of interest well (as shown in Fig. 1(c-d)). A Kagome lattice (Fig. 3a) has three sites (1, 2 and 3) arranged in a triangular topology, and each of these three sites is in turn laid out in a 2D hexagonal Bravais lattice. The tight-binding (TB) Hamiltonian for the Kagome lattice is as follows:

$$H = \sum_i \varepsilon_i c_i^+ c_i + \sum_{<ij, i \neq j>} t_{ij} c_i^+ c_j \qquad (1)$$

where $\varepsilon_i$ is the on-site energy at site $i$, $c_i^+$ and $c_i$ are the creation and annihilation operators at site $i$ respectively, and $t_{ij}$ parameters are the hopping terms where we consider only nearest-neighbor interactions for simplicity. In a typical Kagome lattice, the on-site energies are equal for all sites, and similarly, all hopping terms are equal. In that case, the band structure for the Kagome lattice has a flat band that is degenerate with dispersive bands, as shown in Fig. 3a. There is also a Dirac cone feature where the two dispersive bands cross. Any effect (such as spin-orbit coupling) which breaks the degeneracy between the flat band and the dispersive bands will also add some dispersion to the flat band.[20]

How can one obtain an isolated flat band as we have in 2D-CAP? We can obtain an isolated flat band by changing the parameters in the Kagome lattice to introduce asymmetry in the lattice. When the hopping parameter $t'$ is reduced (to a non-zero value) relative to $t$ (see Fig. 3a), while keeping other parameters the same, we see that the two dispersive bands now avoid each other, and the originally flat band from the Kagome lattice becomes dispersive (Fig. 3(b-d)). The band with intermediate energy (middle band) starts to become flatter as *t'* decreases. When *t'* = 0, this middle band becomes flat, and at the same time, the other two bands cross with the three bands being degenerate at the M point (Fig. 3e). Thus, one obtains another flat band, but this flat band is still degenerate with other bands. To obtain an isolated flat band, we can tune the energy of the flat band while keeping *t'* = 0, by changing the on-site energy $\varepsilon_3$ at site 3, as shown in Fig. 3(f-i). In this way, one can obtain an isolated flat band using a Kagome-like lattice. This type of flat band has been classified as a non-singular flat band in recent literature.[40]

2D-COFs can be designed from bottom-up, and are also very stable even at room temperatures. As such, they are excellent candidates as real materials for testing the predictions made from simple lattice Hamiltonians. We fit an effective Kagome-like Hamiltonian to the DFT bands of CAP-1 and 2D-CAP by minimizing the following cost function:

$$c(\varepsilon_1, \varepsilon_2, \varepsilon_3, t, t') = \sum_{\vec{k},j} w(\vec{k})(E_j^{TB}(\varepsilon_1, \varepsilon_2, \varepsilon_3, t, t', \vec{k}) - E_j^{DFT}(\vec{k}))^2 \qquad (2)$$

where $w(\vec{k})$ refers to the weight of the k-point in the Brillouin zone. $E_{TB}$ and $E_{DFT}$ represent the TB and DFT Kohn-Sham eigenvalues, respectively. The resulting TB band structures (red dashed lines) match reasonably closely with the DFT bands (Fig. 1(c-d)). Examining the fitting parameters, we see that the hopping terms for CAP-1 are *t* = -0.30 eV and *t'* = -0.16 eV, while those for 2D-CAP are *t* = 0.31 eV and *t'* = 0.00 eV. The parameter *t'* corresponds to hopping between the bridge sites and the Kagome sites on the chains, while *t* is the hopping term between the two Kagome sites on the chains (Fig. 1(a-b)). There is a large asymmetry between *t'* and *t*, especially for 2D-CAP where *t'* = 0.00 eV, corresponding to the flat band scenario in Fig. 3(e-i). From this information, we can deduce that one approach to designing COFs with flat bands is to utilize 'I'-shaped motifs and 'T'-shaped linkers, in order to obtain asymmetric hopping parameters in the underlying effective lattice. In contrast to the case of a '+'-shaped linker, the 'T'-shaped linker leads to a physical staggered arrangement of the bridge units, which helps to reduce the value of *t'*.

Furthermore, it is evident that the longer bridge unit in 2D-CAP compared to CAP-1 is also important for reducing $t'$. The presence of nitrogen atoms on the bridge sites but not the chain further helps to decouple the electronic states on the bridge sites from those of the chain sites, reducing $t'$. These nitrogen atoms also serve to change the on-site energies at the bridge sites, thus isolating the flat band from the dispersive bands (Fig. 3i). Besides isolated flat bands, the physical arrangement of chains and bridge sites in these COFs also allows for the existence of physics of mixed dimensions in a single 2D material system, leading to a relatively unexplored research frontier.

In the following, we use 2D-CAP as a case study to explore some electronic properties of a 2D material with an isolated flat band and physics of mixed dimensions. Ultrathin sheets of 2D-CAP had been exfoliated from the bulk layered material in previous work.[27] The phonon band structure for the monolayer has no imaginary modes (Fig. S2), showing that the monolayer is dynamically stable.

The large electron-hole asymmetry naturally leads to a large Seebeck coefficient with a peak value of ~1500 µV/K computed at 300K, at very low carrier concentrations (see Fig. S3). The large peak value of the Seebeck coefficient is comparable to those of known high performance thermoelectric materials.[41-43] Due to the quasi-1D character of the VBM wave function (Fig. 2c), the effective mass for the VBM at $\Gamma$ along $q_{bridge}$ (~1.13$m_0$) is ~ 6 times larger than that along $q_{chain}$ (~0.19$m_0$). Because the CBM is very flat, the electron-hole asymmetries in both the 'bridge' and 'chain' directions are very large, and the difference in Seebeck coefficients computed for the 'bridge' and 'chain' directions is relatively small (Fig. S3). However, the electronic conductivity is inversely proportional to the effective mass, and will be significantly smaller along the 'bridge' direction, which can result in a large thermoelectric figure of merit.

We also compute the quasiparticle band structure of 2D-CAP using the GW approximation (Fig. 4a). The essential features of the DFT band structure discussed above are still present in the GW band structure, except that the GW band gap is 3.44 eV, compared to a band gap of 1.16 eV obtained from DFT. Besides large electron-hole asymmetry, the mixed dimensionality in 2D-CAP also shows up in the optical properties of 2D-CAP, which we compute using GW-BSE calculations (solid lines; Fig. 4b). 2D-CAP exhibits linear dichroism, with negligible light absorption when the incident light is polarized perpendicular to the chains. The linear dichroism results from the depolarization effect perpendicular to the chains.[3] The low dimensionality of 2D-CAP results in large exciton binding energies (> 1 eV), which are evident from the large red-shift of the GW-BSE optical absorption spectra compared to the spectra obtained without considering electron-hole interactions (dashed lines; Fig. 4b). The lowest energy bright excitons are within the visible range. Exciton I (the lowest energy bright exciton) has relatively smaller oscillator strength compared to Exciton II. Exciton II involves a transition between VBM and CBM+1 (Fig. 4a), and the exciton wave function is distributed along the chain (Fig. 4d). On the other hand, Exciton I involves a transition from VBM to CBM (Fig. 4a), which is the flat band arising from the bridge sites. The relatively weak oscillator strength comes from the fact that the VBM has only a small weight on the bridge sites (Fig. 2c). It is interesting that absorption of light polarized parallel to the chains can result in a 2D exciton that is distributed across multiple bridge sites in the 2D material (Fig. 4c). These observations are a manifestation of the physics of mixed dimensionalities present in this system.

The isolated flat band and quasi-1D VBM character suggests that 2D-CAP and similar systems are interesting platforms to study the physics of flat bands and mixed dimensionalities. Hole-doping the system can lead to quasi-1D physics. Further adjustments to the design of the COF structure, such as an increase in length of the bridges and a further increase in the on-site energies of the bridge sites relative to the chain sites, may lead to a VBM wave function that has minimal weight on the bridge sites. It would be interesting to see if exotic 1D physics can be realized when such systems are hole-doped. On the other hand, electron-doping of 2D-CAP can result in an array of interesting quantum phases related to the isolated flat band. The bandwidth W of the 'isolated flat band' in 2D-CAP is 28 meV, which is comparable to that of twisted bilayer $MoS_2$[7] and twisted bilayer graphene near the magic angle.[44] DFT calculations predict that doping 2D-CAP with 0.2 electrons per unit cell leads to a breaking of spin degeneracies (Fig. S4). Similar results are obtained for doping concentrations of between 0 and 2 electrons per unit cell. We also estimate the on-site Coulomb interaction $U^7$ at the bridge sites to be 730 meV (see Supplemental Information). The fact that U >> W indicates that 2D-CAP is a potential candidate for realizing the Mott-insulating phase at half-filling of the flat band.

**Conclusions**

In this work, we demonstrate that a recently synthesized 2D COF (2D-CAP) hosts the long sought-after isolated flat band, which can lead to strongly correlated physics upon electron-doping. 2D CAP also provides a platform for the study of physics of mixed dimensions (0D, 1D and 2D) within a single 2D sheet, without the need for introducing defects or adsorbates. By comparing 2D-CAP with another similar COF, we elucidate strategies for the rational design of 2D COFs which have isolated flat bands, as well as physics of mixed dimensions. The key design strategies are the use of 'I'-shaped building blocks and 'T'-shaped linkers, as well as the choices for the length and chemical nature of the bridge and chain units of the resulting COF. Using 2D-CAP as an example, we show how mixed dimensional physics can result in large Seebeck coefficients due to strong electron-hole asymmetries, as well as linear dichroism in conjunction with unique strongly bound 2D excitons spread over the quasi-0D bridge sites. Further optimization of the structural and electronic properties within this family of 2D COFs may result in other interesting phenomena, such as a gate-controlled switching between strongly-correlated quantum phases arising from the isolated flat band, and exotic phases associated with quantum-confined 1D systems.

**Conflicts of Interest**

There are no conflicts of interest to declare.

**Acknowledgements**

We acknowledge support from the Singapore National Research Foundation, Prime Minister's Office, under its medium-sized centre program. J.W. acknowledges a graduate student scholarship from NUS. Computations were performed on the NUS Graphene Research Centre cluster and National Supercomputing Centre Singapore (NSCC).

**Figures:**

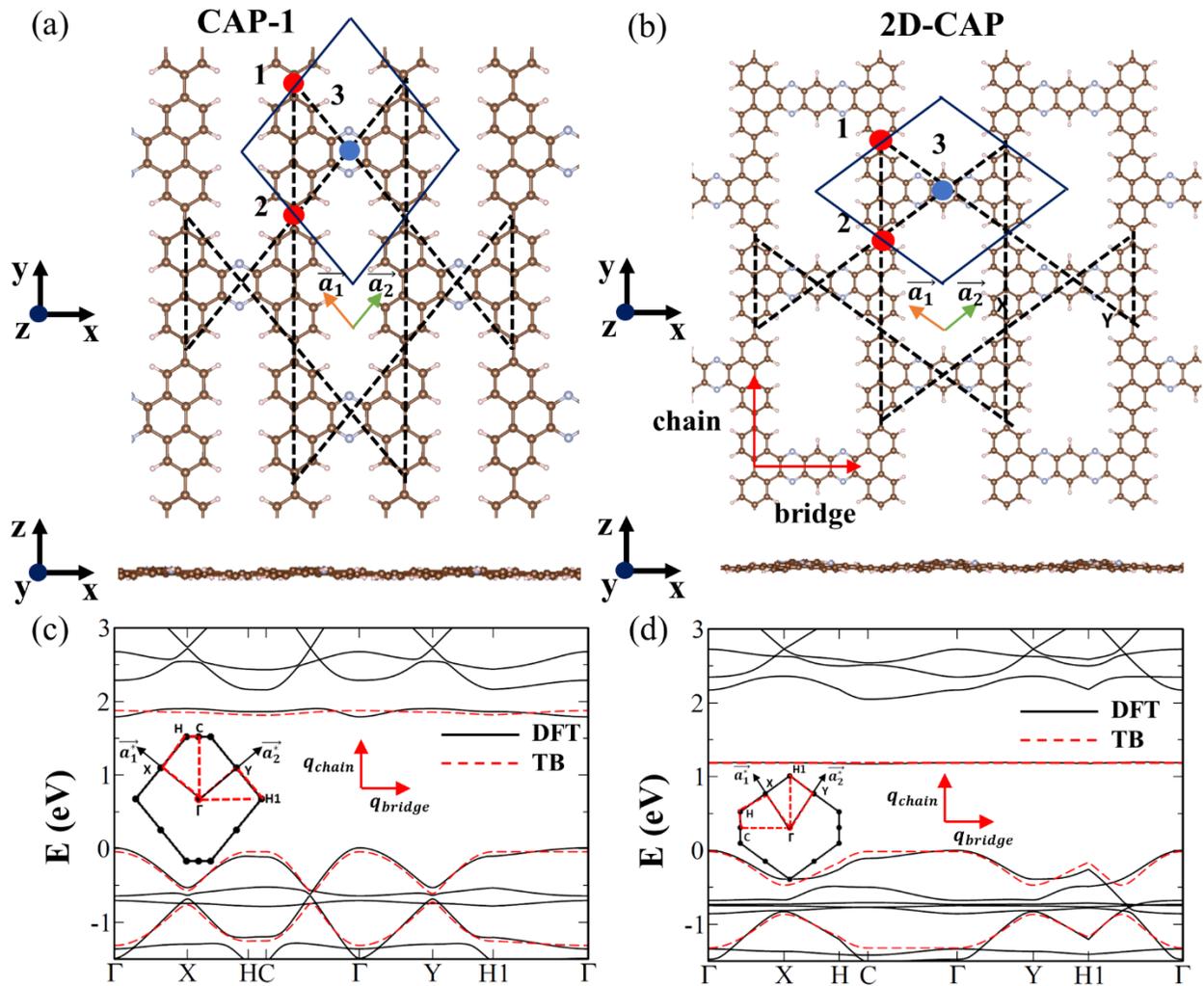

**Figure 1.** Atomic structure and band structures of 2D COFs. (a-b) Atomic structures of (a) CAP-1 and (b) 2D-CAP. The unit cells are indicated by the purple parallelograms. (c-d) Band structures of (c) CAP-1 and (d) 2D-CAP. The tight-binding (TB) models are obtained by fitting the DFT band structures to Kagome-like lattices as indicated in (a) and (b) by dashed black lines. The TB parameters are indicated in Table 1. The high symmetry directions for the band structures are shown in the inset in (c) and (d). The bridge sites of 2D-CAP couple only weakly to the chains and to one another, forming nearly zero dimensional components arranged in a stable 2D lattice. A flat band that is isolated in energy from other bands arises.

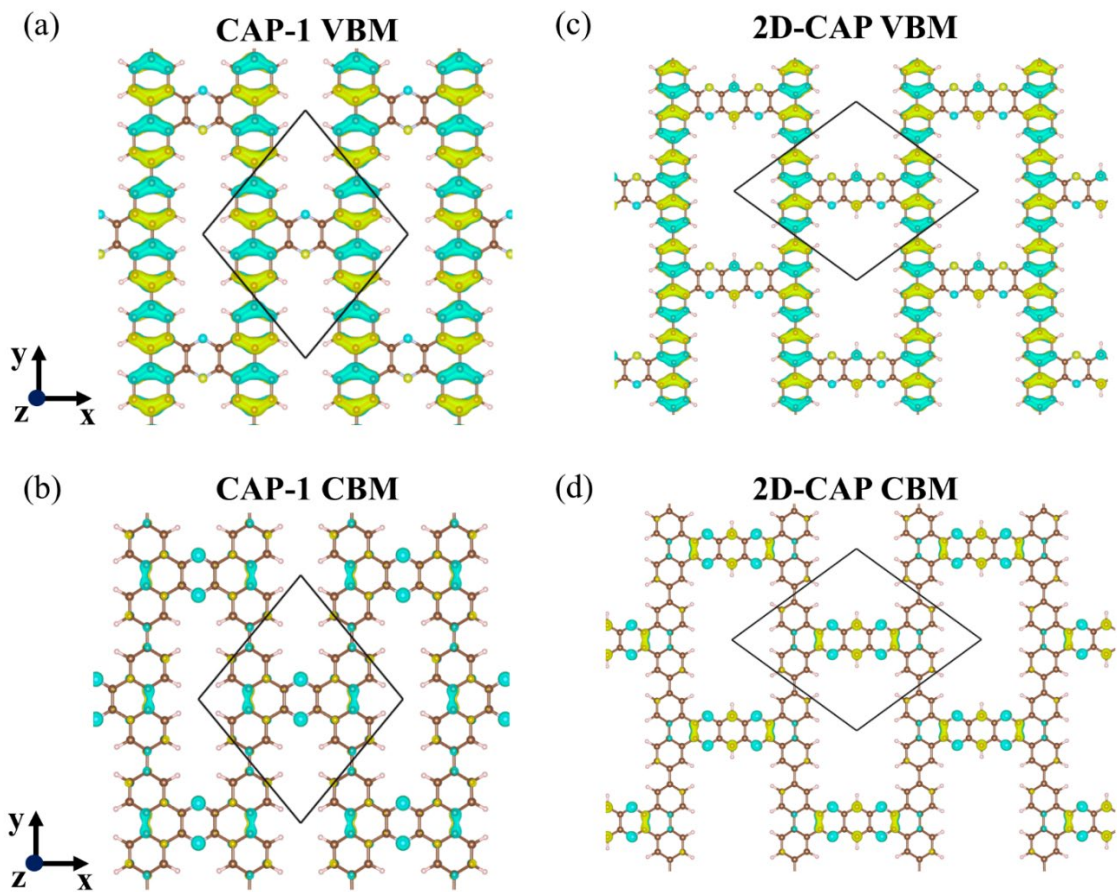

**Figure 2.** Gamma point Kohn-Sham wave functions for (a) VBM and (b) CBM of CAP-1, and (c) VBM and (d) CBM of 2D-CAP. Isosurface values are chosen to be 10 % of the maximum value in each case. Positive and negative values are indicated by green and yellow, respectively.

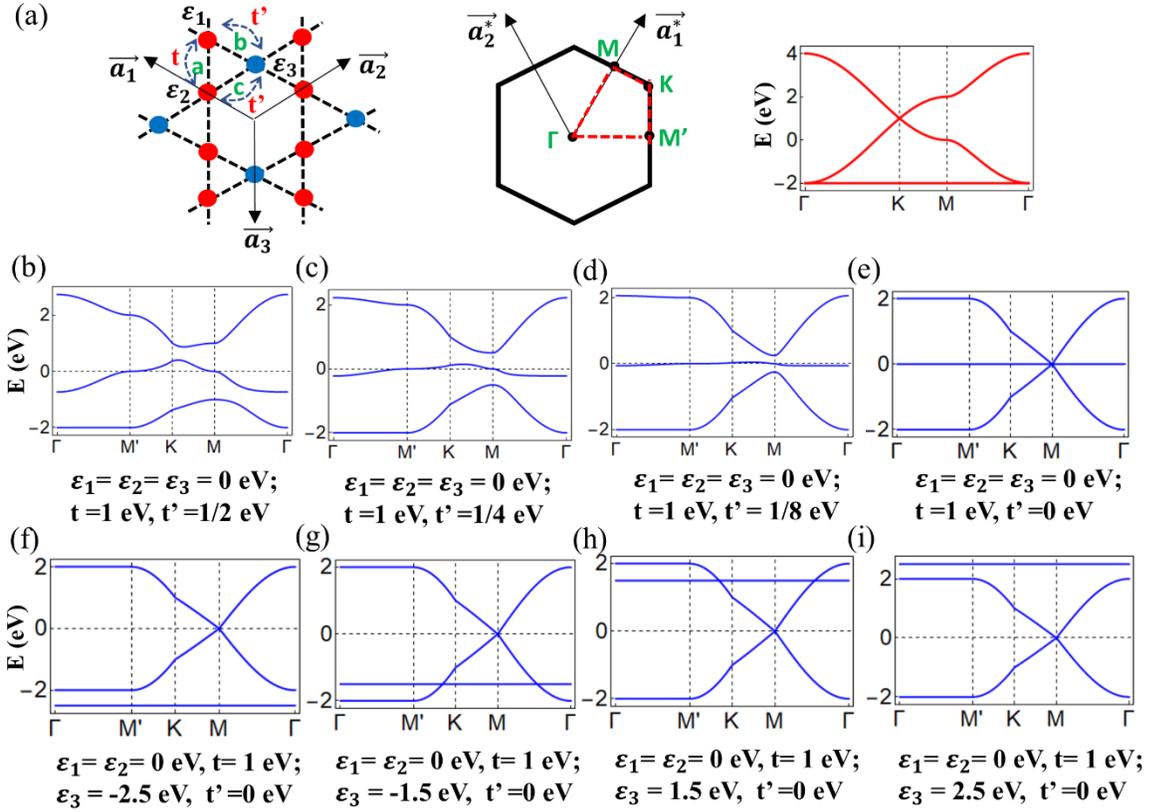

**Figure 3.** Isolated flat band from Kagome-like lattice. (a) Real space lattice, Brillouin zone and band structure of a Kagome lattice ($\varepsilon_1 = \varepsilon_2 = \varepsilon_3 = 0$ eV, t' = t = 1 eV), (b-e) Band structures of Kagome-like lattices with $\varepsilon_1 = \varepsilon_2 = \varepsilon_3 = 0$ eV, t' < t, (f-i) Band structures of Kagome-like lattices with t' = 0 eV, $\varepsilon_1 = \varepsilon_2 = 0$ eV, and $\varepsilon_3 \neq 0$ eV.

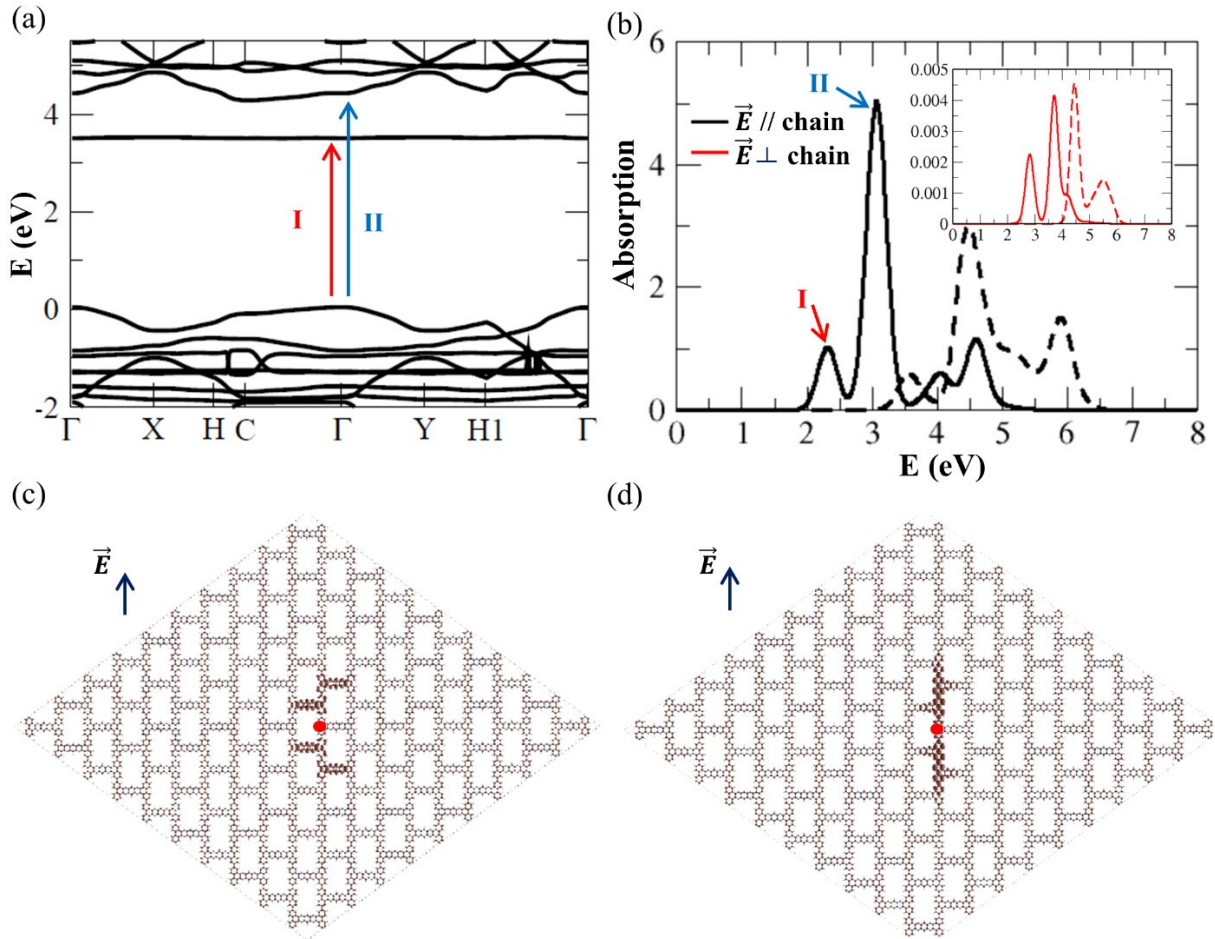

**Figure 4.** Linear dichroism from quasi-1D nature of chains in 2D-CAP. (a) Quasiparticle (GW) band structure and (b) optical absorption spectra, including the effect of electron-hole interactions, in 2D-CAP. The optical absorption spectra without electron-hole interactions is shown in dashed lines. The optical absorption spectrum when the direction of light polarization $\vec{E}$ is perpendicular to the chain is shown as inset. Visualization of exciton wave functions for (c) exciton I and (d) exciton II as marked in (b). The red dot represents the position of a hole and the electron distribution is shown, and the corresponding single particle transitions are indicated in (a). The direction of light polarization $\vec{E}$ is indicated by blue arrows in (c) and (d). There is negligible absorption of light polarized perpendicular to the chains due to the quasi-1D nature of the chains.

**Table 1.** Optimized tight-binding parameters (in eV) for monolayer CAP-1 and 2D-CAP.

| system | $\varepsilon_1$ | $\varepsilon_2$ | $\varepsilon_3$ | $t$ | $t'$ |
|---|---|---|---|---|---|
| CAP-1 | -0.73 | -0.57 | 1.81 | -0.30 | -0.16 |
| 2D-CAP | -0.87 | -0.47 | 1.19 | 0.31 | 0.00 |

# Supplementary Information

# Isolated Flat Bands and Physics of Mixed Dimensions in a 2D Covalent Organic Framework

*Juefan Wang[1,2], Su Ying Quek[1,2, *]*

[1]Department of Physics, National University of Singapore, 2 Science Drive 3, 117551, Singapore

[2]Centre for Advanced 2D Materials, National University of Singapore, Block S14, Level 6, 6 Science Drive 2, 117546, Singapore

**Methods**

The thermoelectric properties of 2D-CAP were studied based on DFT and Boltzmann transport theory [1]. DFT calculations were performed within the projector augmented-wave formalism [2] as implemented in the Vienna ab initio simulation package (VASP) [3]. The generalized gradient approximation [4] was used for the exchange-correlation functional. A plane-wave basis set with a kinetic energy cutoff of 550 eV was employed. The Brillouin zone was sampled on a 4 × 4 × 1 Monkhorst-pack grid for the self-consistent field calculation and a denser k grid of 20 × 20 × 1 for the non-self-consistent field calculation to obtain the Kohn-Sham energies for computing the Seebeck coefficients. Seebeck coefficients were calculated by means of Boltzmann semiclassical theory, within the constant relaxation time approximation, using the BoltzTraP2 code [5].

The on-site Coulomb interaction U at the bridge sites is estimated following Ref. [6]. The Coulomb interaction U is given by $U=\frac{e^2}{4\pi\varepsilon\epsilon_0 d}$, where the in-plane dielectric constant of 2D-CAP, $\varepsilon$, is computed from density-functional-perturbation-theory, with a value of 3.1. d is estimated by first obtaining the macroscopic planar-average of the localized wavefunction charge density $|\psi(r)|^2$, and d is the full-width-half-maximum of this function. The value of d is 6.4 Å. Our computed U = 730 meV.

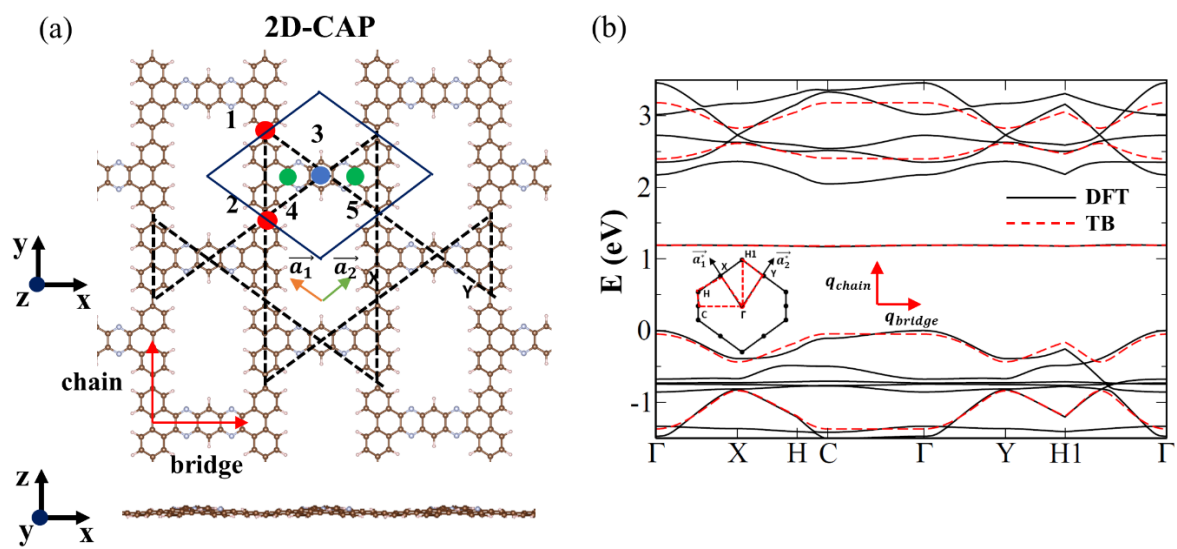

**Figure S1.** (a) The Kagome-honeycomb tight-binding (TB) model for monolayer 2D-CAP. The three Kagome sites are labelled as 1, 2 and 3, respectively. The two honeycomb sites are labelled as 4 and 5, respectively. (b) Band structure of 2D-CAP from DFT (black solid line) and TB model (red dashed line). The fitting parameters are shown in Table S1.

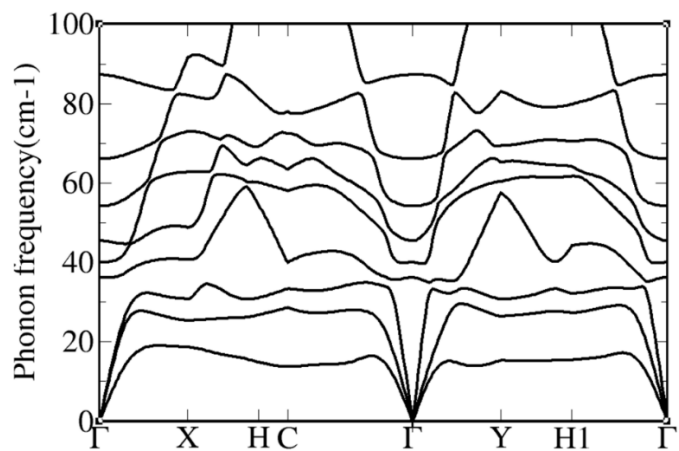

**Figure S2.** Phonon band structure of monolayer 2D-CAP.

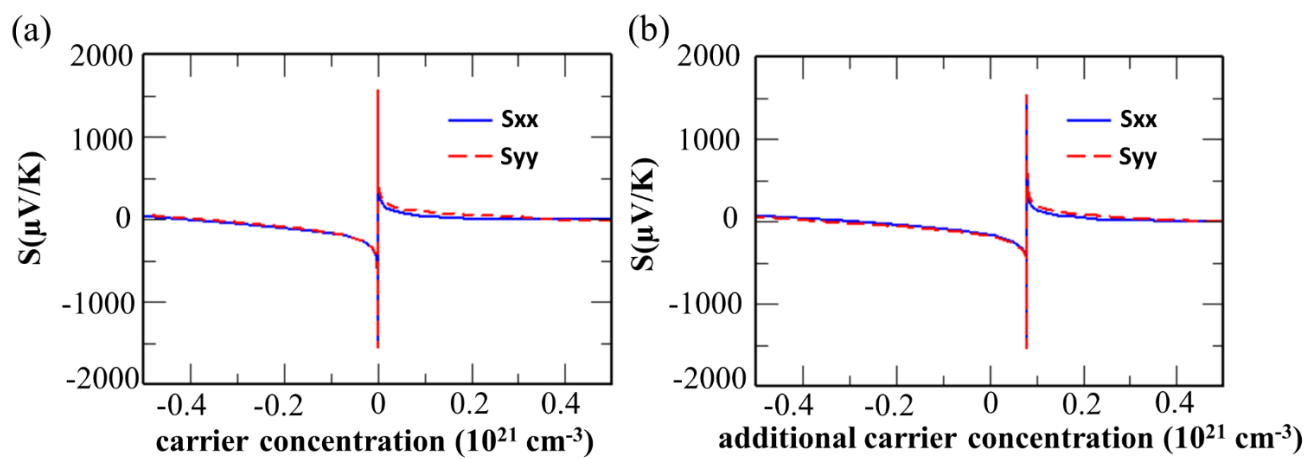

**Figure S3.** Seebeck coefficients (μV/K) computed at 300 K, as a function of carrier concentration ($10^{21}$ cm$^{-3}$) of (a) undoped 2D-CAP (b) 2D-CAP doped with 0.2 e$^-$. The simulation in (b) illustrates that spin polarization does not have a significant effect on the Seebeck coefficients, and that the rigid band approximation, used in (a), works reasonably well. Directions x and y correspond to the 'bridge' and 'chain' directions, respectively. $S_{yy}$ is slightly larger than $S_{xx}$ because of the asymmetry in the band dispersions in the valence band.

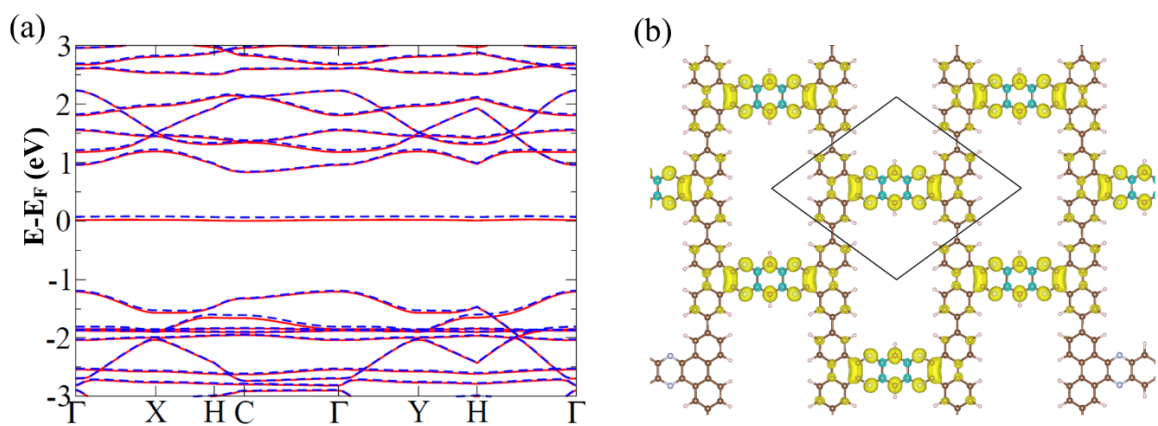

**Figure S4.** Electronic structure of 2D-CAP doped with 0.2 electrons. (a) Spin-polarized band structure. Red solid lines: majority spin, blue dashed lines: minority spin. (b) The corresponding spin density plot. The majority spin and minority spin density are indicated by yellow and green, respectively. The isosurface value is 5% of the maximum.

**Table S1.** Optimized tight-binding parameters (in eV) for monolayer 2D-CAP using a Kagome-honeycomb lattice model. The on-site energies of each site are indicated by $\varepsilon_i$ (i from 1 to 5). For the hopping terms, we use the following notations - $t_1 = t_{12}(t_{21})$: hopping between Kagome site 1 and site 2; $t_2 = t_{13}(t_{31}) = t_{23}(t_{32})$: hopping between Kagome site 1/2 and Kagome site 3; $t_3 = t_{14}(t_{41}) = t_{15}(t_{51}) = t_{24}(t_{42}) = t_{25}(t_{52})$: hopping between Kagome site 1/2 and honeycomb site 4/5; $t_4 = t_{34}(t_{43}) = t_{35}(t_{53})$: hopping between Kagome site 3 and honeycomb site 4/5; $t_5 = t_{45}(t_{54})$: hopping between honeycomb site 4 and honeycomb site 5.

| on-site energy(eV) | $\varepsilon_1$ | $\varepsilon_2$ | $\varepsilon_3$ | $\varepsilon_4$ | $\varepsilon_5$ |
|---|---|---|---|---|---|
| | -0.55 | -0.11 | 1.20 | 2.29 | 2.52 |
| hopping term (eV) | $t_1$ | $t_2$ | $t_3$ | $t_4$ | $t_5$ |
| | 0.51 | 0.00 | 0.68 | 0.10 | 0.00 |